# Il progetto *Lab2Go* per la diffusione della pratica laboratoriale nelle Scuole Secondarie di II grado

Mirco Andreotti[1], Pia Astone[2], Donatella Campana[3], Antonella Cartoni[4], Fausto Casaburo[5,2,*], Francesca Cavanna[6], Gianluigi Cibinetto[1], Antonella Dalla Cort[4], Giulia De Bonis[2], Marta Della Seta[7], Francesca Di Mauro[8], Giuseppe Di Sciascio[9], Riccardo Faccini[5,2], Federica Favino[10], Luca Iocchi[11], Marcello Lissia[12], Giulia Morganti[8], Mauro Mancini[2], Giovanni Organtini[5,2], Francesco Pennazio[6], Francesco Piacentini[5,2], Alina Piras[8], Maria Ragosta[13,3], Lorenzo Roberti[5,14], Anna Rita Rossi[15], Laura Sadori[16], Francesco Safai Tehrani[2], Stefano Sarti[5,2]

[1] *Istituto Nazionale di Fisica Nucleare (INFN)- Sezione Ferrara*

[2] *Istituto Nazionale di Fisica Nucleare (INFN)- Sezione Roma*

[3] *Istituto Nazionale di Fisica Nucleare (INFN)- Sezione Napoli*

[4] *Sapienza Università di Roma- Dipartimento di Chimica*

[5] *Sapienza Università di Roma- Dipartimento di Fisica*

[6] *Istituto Nazionale di Fisica Nucleare (INFN)- Sezione Torino*

[7] *Sapienza Università di Roma- Dipartimento di Scienze della Terra*

[8]*IIS Anagni, Sede Liceo Scientifico "D. Alighieri" di Fiuggi (FR)*

[9] *Istituto Nazionale di Fisica Nucleare (INFN)- Sezione Tor Vergata*

[10] *Sapienza Università di Roma- Dipartimento di Storia Antropologia Religioni Arte e Spettacolo (SARAS)*

[11] *Sapienza Università di Roma- Dipartimento di Ingegneria Informatica Automatica e Gestionale*

[12] *Istituto Nazionale di Fisica Nucleare (INFN)- Sezione Cagliari*

[13] *Scuola di Ingegneria – Università della Basilicata*

[14] *Istituto Nazionale di AstroFisica (INAF)- Osservatorio astronomico di Roma*

[15] *Sapienza Università di Roma- Dipartimento di Biologia e Biotecnologie*

[16] *Sapienza Università di Roma- Dipartimento di Biologia Ambientale*

[*] *Corresponding author*

ABSTRACT

Even if laboratory practice is essential for all scientific branches of knowledge, it is often neglected at High School, due to lack of time and/or resources. To establish a closer contact between school and experimental sciences, the University Sapienza of Roma and the Istituto Nazionale di Fisica Nucleare (INFN) launched the *Lab2Go* project, with the goal of spreading laboratory practice among students and teachers in high schools.

**Introduzione**

Uno dei primi argomenti affrontati da ogni studente di scuola secondaria di II grado, nelle ore dedicate alle materie scientifiche, è il *metodo scientifico*. Fin dai primi approcci, lo studente impara che alla base di tali discipline c'è l'osservazione dei fenomeni, la formulazione di un modello e la sua (eventuale) conferma sperimentale. L'esperimento (sia nella fase di "osservazione" che in quella di "conferma") è, dunque, un punto cardine di ogni ambito scientifico; tuttavia, complici i laboratori non sempre attrezzati (o a volte addirittura assenti) e la necessità degli insegnanti di preparare gli studenti all'Esame di Stato, la parte sperimentale è talvolta trascurata nelle scuole. Tutto ciò può comportare un disinteresse degli studenti per le discipline scientifiche che, ai loro occhi, appaiono solamente come qualcosa di astratto e il cui unico scopo è quello di risolvere gli esercizi assegnati dal docente. Per correggere questa percezione e avvicinare gli studenti alla pratica laboratoriale, a partire dall'a.s. 2016/17 l'Università Sapienza di Roma, in collaborazione con la Sezione di Roma dell'Istituto Nazionale di Fisica Nucleare (INFN), ha avviato il progetto Lab2Go, finanziato dal *Piano Lauree Scientifiche* (PLS) nell'ambito delle attività di *Alternanza Scuola Lavoro* (ASL), oggi *Progetti per le Competenze Trasversali e l'Orientamento* (PCTO). Durante la loro partecipazione a questo progetto, denominato Lab2Go, gli studenti non solo hanno la possibilità di effettuare esperimenti sfruttando gli strumenti presenti nel loro laboratorio scolastico, ma hanno anche il compito di catalogare tale strumentazione e le esperienze svolte sulla WiKi di Lab2Go, ospitata nell'infrastruttura informatica fornita dalla citata sezione INFN di Roma [1-2]. Le schede didattiche catalogate, così come tutte le informazioni riportate sulle pagine WiKi, possono quindi essere utilizzate liberamente, non solo dalle altre scuole partecipanti al progetto, ma da chiunque abbia interesse a ripetere quell'esperimento o avere informazioni su quell'argomento. Inoltre, grazie alla catalogazione della strumentazione del laboratorio scolastico, è possibile sapere quali strumenti possiede ciascuna scuola e attivare un processo di condivisione degli stessi tra scuole limitrofe. L'idea del progetto è quindi di creare un vero e proprio *laboratory sharing* (di strumenti, esperienze e pratiche didattiche). Lab2Go è, dunque, un progetto che si ripropone di creare una rete di scuole che condividono pratiche laboratoriali con l'ausilio di Università ed Enti di Ricerca. In questo modo, si intende diffondere la pratica laboratoriale sia tra le scuole che all'interno delle singole scuole, fornendo a tutti i docenti materiale facilmente consultabile sull'utilizzo dei laboratori didattici all'interno dei propri Istituti. A tal fine sono stati organizzati anche corsi di formazione per i docenti all'interno dei laboratori scolastici stessi. [3-6].

1. **Il progetto *Lab2Go***

   Al progetto *Lab2Go* possono partecipare tutte le scuole d'Italia che ne fanno richiesta, compilando l'apposito modulo di adesione che, nell'estate che precede ogni anno scolastico, viene pubblicato sul sito web.infn.it/lab2go. Gli scopi del progetto sono:
   - Catalogazione degli strumenti e delle dotazioni presenti nei laboratori scolastici;
   - Esecuzione di esperimenti in laboratorio;
   - Condivisione di esperienze e creazione di "reti" di scuole (anche attraverso la pubblicazione di contenuti sulle pagine Wiki);
   - Aggiornamento dei docenti riguardo la realizzazione di attività sperimentali.

Tra i mesi di novembre-dicembre di ogni anno, viene generalmente organizzato un evento iniziale a cui partecipano tutte le scuole aderenti al progetto e, durante il quale, gli organizzatori ne illustrano gli scopi principali.

A ogni scuola sono assegnati:
- un *Tutor esterno*: un docente universitario o (per fisica) un ricercatore INFN che ha aderito al progetto;
- un *Borsista*: un laureando magistrale o un dottorando in una delle discipline scientifiche aderenti al progetto (a cui viene attribuita una borsa di tutoraggio, ex Legge 170/2003, messa a bando dall'Università Sapienza di Roma, dall'INFN o attraverso i fondi PLS);
- un *Tutor interno*: un docente della scuola che collabora col tutor esterno e col borsista nello svolgimento delle attività.

Il tutor esterno e il borsista organizzano, in accordo con il tutor interno, una serie di incontri da svolgersi nel laboratorio scolastico, durante i quali gli studenti eseguono esperimenti (Fig. 1), imparano la sintassi Dokuwiki [7] e riportano su WiKi l'attività svolta.

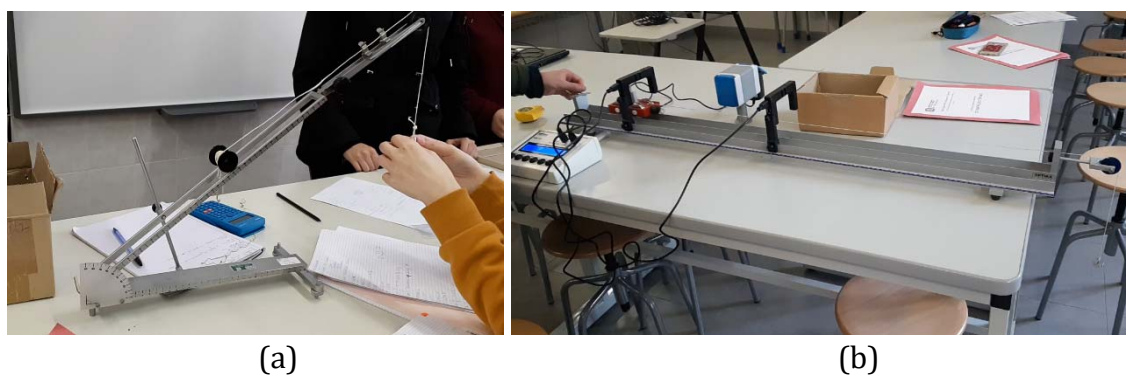

(a)  (b)

Figura 1: Studenti nella preparazione degli esperimenti. (a) Equilibrio di un corpo su un piano inclinato, presso il Liceo Scientifico "*D. Alighieri*" di Anagni (FR)- sezione staccata di Fiuggi (FR) a.s. 2019/20.
(b) Moto uniformemente accelerato con l'uso di rotaia a basso attrito presso il Liceo Classico "*P. Albertelli*" di Roma, a.s. 2019/20.

A questi incontri, inoltre, si possono affiancare ulteriori incontri nei laboratori universitari, dedicati alla compilazione della documentazione, o ad attività sperimentali o di orientamento. Infine, a conclusione dell'anno scolastico, si svolge un evento finale che vede la partecipazione di tutte le scuole aderenti e durante il quale gli studenti possono presentare l'attività svolta nell'ambito del progetto (Fig. 2).

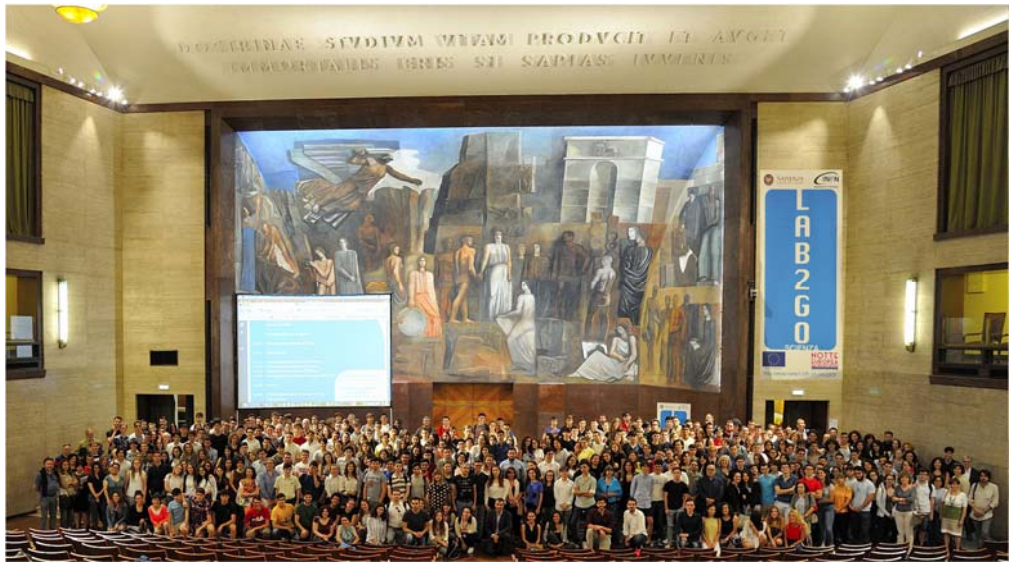

Figura 2: Evento finale Lab2go a.s. 2017/18.

### 1.1 La diffusione del progetto *Lab2Go*

Quando è iniziato, *Lab2Go* era esclusivamente dedicato alla Fisica e ad esso partecipavano solamente scuole del Lazio (prevalentemente della provincia di Roma). Grazie all'impegno dei responsabili, in pochi anni, il progetto si è rapidamente sviluppato abbracciando ulteriori discipline:

1. Biologia Animale: si occupa della catalogazione delle preparazioni istologiche e dei reperti animali presenti nelle scuole e della messa a punto di osservazioni ed esperimenti con materiali biologici a fresco, corredati dall'allestimento di schede operative.
2. Botanica: si occupa del verde dell'Istituto scolastico con attività e strumenti diversi, che mirano sia alla riqualificazione di aree preesistenti con l'inserimento di specie arboree e arbustive autoctone che alla realizzazione di orti urbani con piante ortive locali.
3. Chimica: si occupa della catalogazione del materiale presente e della realizzazione di semplici esperienze nei laboratori scolastici (Fig. 3).
4. Scienze della Terra: si occupa del censimento della collezione di rocce eventualmente presente a scuola, della selezione di campioni di cui effettuare sezioni e dello studio dei terremoti.
5. Robotica: si occupa della costruzione hardware e della programmazione software di robot quale ad esempio *MARRtino*, un Robot *Arduino- based open source* e *open hardware*, sviluppato dal Dipartimento di Ingegneria Informatica Automatica e Gestionale dell'Università Sapienza di Roma [8].
6. Musei Scientifici: si occupa di catalogare le collezioni di strumenti di interesse storico-scientifico, conservate presso gli istituti scolastici e d'indagarne l'origine e l'evoluzione nel tempo (Fig. 4).

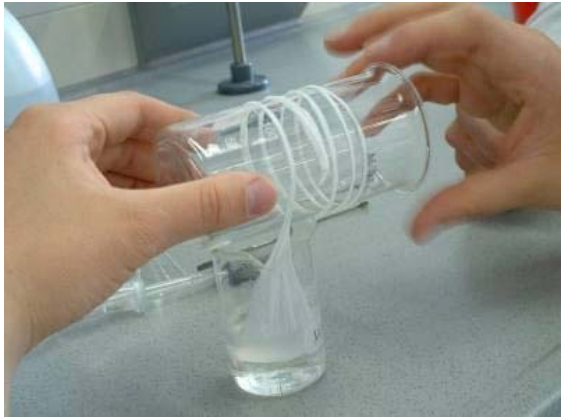 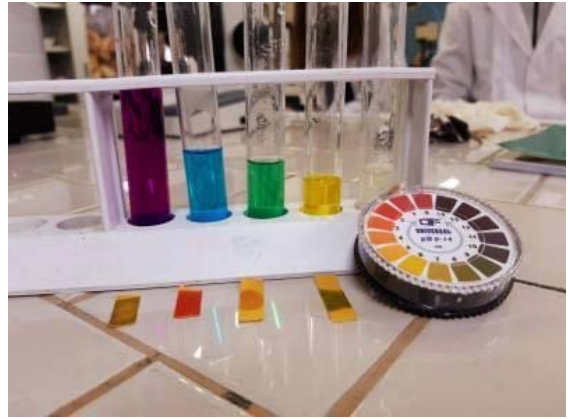

(a) (b)

Figura 3: (a) Sintesi del Nylon, una poliammide ottenuta per reazione tra una diammina ed un acido bicarbossilico presso il Liceo Scientifico "*G. Sulpicio*" di Veroli (FR), a.s. 2018/19. (b) Soluzioni di diversi sali inorganici colorati che cambiano il pH dell'acqua in diverso modo da molto acida (cartina tornasole rossa) a molto basica (cartina tornasole blu) presso l'ISISS "*G. De Sanctis*" di Roma, a.s. 2017/18.

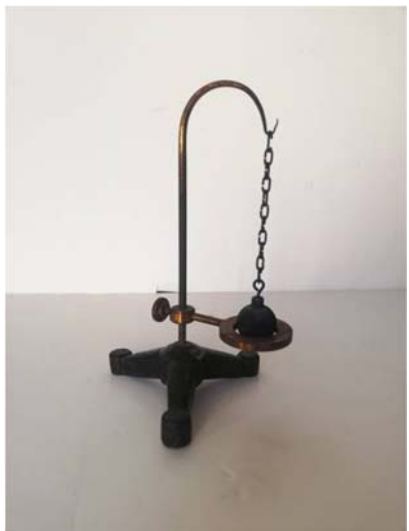 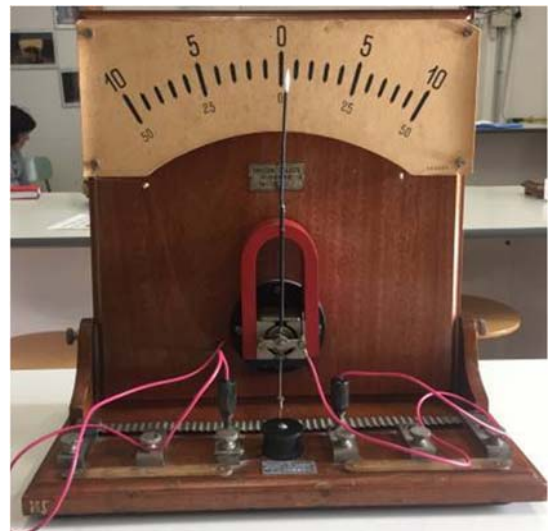

(a) (b)

Figura 4: Esempi di antichi strumenti catalogati nella sezione "Musei Scientifici" della Wiki Lab2Go. (a) Un anello di Gravesande conservato presso il Liceo Classico "*T. Tasso*" di Roma. (b) Un galvanometro conservato presso il Liceo Classico "*G. Cesare*" di Roma.

La diffusione del progetto non ha riguardato solamente le ulteriori discipline che si sono aggiunte alla Fisica con il crescente coinvolgimento dell'Università Sapienza di Roma, ma è stata anche geografica (Fig. 5); infatti, l'interesse per il progetto è stato tale da vedere l'adesione di ulteriori sezioni INFN (Roma Tor Vergata, Torino, Bari, Napoli, Cagliari, Ferrara e, dall'a.s. 2021/22, anche Bologna, Laboratori Nazionali del Gran Sasso,

Milano, Padova, Pavia, e Perugia) che hanno iniziato a collaborare al progetto e la partecipazione, per la Fisica, di numerose scuole da Nord a Sud d'Italia.

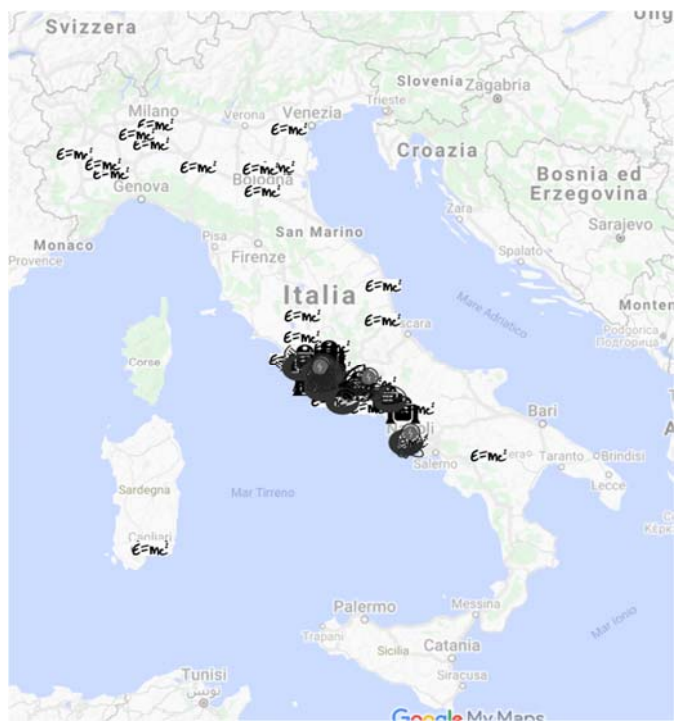

Figura 5: Localizzazione su *Google Maps* delle scuole aderenti al progetto *Lab2Go* 2021/22.

### 1.2 L'epidemia COVID-19 e la nascita di *Lab2Go@home*

Tra la fine di febbraio e gli inizi di marzo 2020, l'Italia è stata colpita da una delle più grandi pandemie della storia: l'emergenza COVID-19. La chiusura delle scuole ha comportato anche l'impossibilità di proseguire gli incontri del progetto *Lab2Go* che, fino a quel momento, aveva sempre fatto uso dei laboratori scolastici e universitari. Quello che poteva essere un problema per il proseguimento delle attività, si è invece trasformato in una proposta e una nuova opportunità che il progetto *Lab2Go* ha fornito agli studenti. In analogia al nome del D.P.C.M. #IO RESTO A CASA, col quale si vietava ogni forma di assembramento su tutto il territorio nazionale, possiamo dire che l'idea di *Lab2Go* è stata #IO SPERIMENTO A CASA. L'emergenza COVID-19, infatti, ha avuto un forte impatto sull'insegnamento sia scolastico che universitario, rendendo necessario il ricorso alla didattica a distanza. Senza dubbio, i disagi legati alla didattica a distanza hanno riguardato in particolar modo le materie scientifiche a causa dell'impossibilità per docenti e studenti di recarsi in laboratorio. Al fine di garantire il prosieguo della pratica laboratoriale anche in questo difficile periodo, si è reso necessario proporre agli studenti esperimenti che potessero essere svolti anche a casa con l'ausilio di materiale a basso costo e facilmente reperibile [9]. Durante tale periodo, dunque, *Lab2Go* ha iniziato una serie di attività on-line, ribattezzate *Lab2Go@home* in cui venivano mostrate la costruzione di strumenti (Fig. 6) e lo svolgimento di esperienze che gli studenti potevano ripetere con materiale facilmente reperibile in casa, pur non disponendo di sofisticata strumentazione, ma tuttavia "centrando" l'obiettivo di Lab2Go di dimostrare che "la scienza passa per le mani" e che la pratica sperimentale è un elemento fondamentale della didattica delle discipline scientifiche.

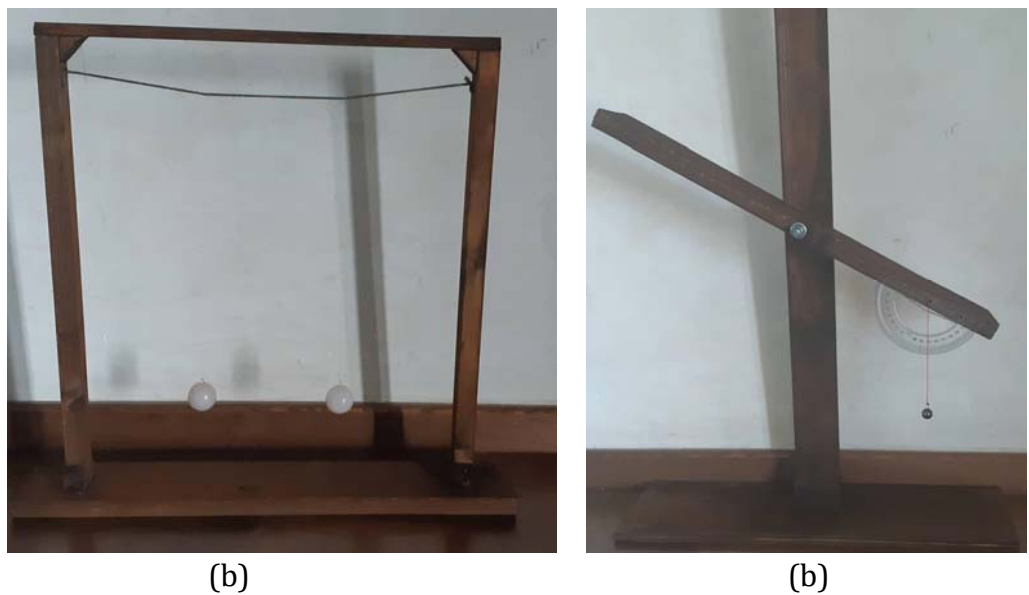

(b) (b)

Figura 6: Esempi di alcuni tra gli strumenti artigianali, costruiti dal borsista Fausto Casaburo e mostrati durante gli incontri di *Lab2Go@Home*. (a) Un sistema di pendoli accoppiati. (b) Un piano inclinato.

Con l'inizio dell'a.s. 2020/21 l'Italia si è ritrovata a far fronte a una nuova ondata di contagi dovuti a COVID-19, così anche quest'anno i numerosi incontri *Lab2Go@home* hanno garantito il proseguo delle attività nelle scuole in cui i laboratori non sono stati frequentabili e, visto l'interesse mostrato dalle scuole per le attività con materiale povero, affiancando il lavoro anche laddove i laboratori erano frequentabili. Al fine di permettere a tutti di rivedere i contenuti trasmessi per la Fisica, gli incontri sono stati registrati e pubblicati sul canale *Youtube,* accessibile dal sito www.youtube.com/channel/UCicLfUmQo5dzedmzjtFh6Ww , *"INFN Edu Physics"*. L'interesse mostrato dalle scuole nei confronti di *Lab2Go@home* ha spinto i responsabili a decidere di proseguire tali attività anche quando l'emergenza COVID-19 sarà terminata.

## 2. Esempi di attività proposte con l'utilizzo di tecnologie e lo sviluppo di competenze di programmazione

### 2.1 L'utilizzo di App per Smartphone

Molto spesso, durante gli incontri di *Lab2Go@home* e gli incontri in presenza laddove i laboratori erano frequentabili, è stato mostrato agli studenti l'utilizzo di app per *smartphone* e in particolare *PhyPhox* (Fig. 7), sviluppata dal Politecnico di Aachen (Germania) e tradotta in italiano da studenti di scuola secondaria di II grado, nell'ambito di un progetto ASL promosso dall'Università Sapienza di Roma sotto la guida del prof. Giovanni Organtini del Dipartimento di Fisica. *PhyPhox* sfrutta i sensori dello *smartphone* per effettuare esperimenti di fisica. Infatti, lo *smartphone* è dotato di numerosi sensori, quali accelerometro, giroscopio, magnetometro e molti altri, che permettono la misura di numerose grandezze fisiche e quindi offrono numerose possibilità di effettuare esperimenti, trasformando il cellulare in un vero e proprio laboratorio a portata di mano [10].

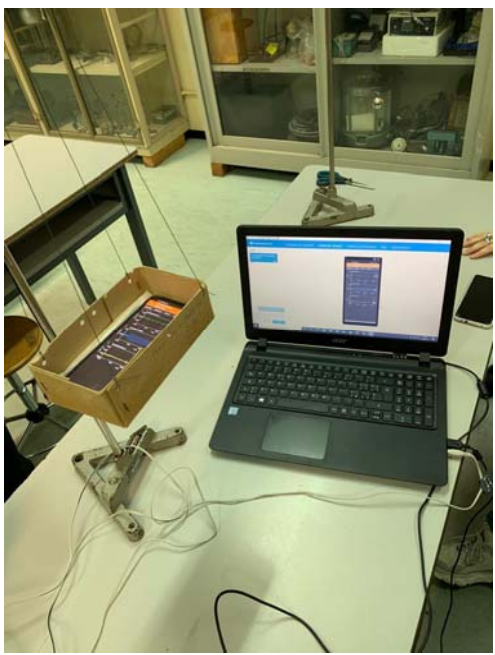

Figura 7: Misura del periodo di oscillazione del pendolo con *PhyPhox* presso il Liceo Scientifico "*L. Trafelli*" di Nettuno (FR), a.s. 2020/21.

Nella disciplina Robotica, lo *smartphone* è usato per guidare il robot tramite comandi vocali (ad esempio, impostare la sequenza di azioni da eseguire durante un esperimento). Nell'ambito del progetto MARRtino è stata anche sviluppata un'app che consente di tradurre comandi vocali in testo (mediante l'uso di *Google Speech API*) e di inviare il testo riconosciuto al robot tramite *Transmission Control Protocol* (TCP). Questo testo (trasformato in una stringa) può essere interpretato dal programma sviluppato dagli studenti (*Python* o *Blockly*) per il controllo del robot.

**2.2 L'utilizzo di Arduino**

La scheda *Arduino* [11] è stata usata nelle attività proposte per la disciplina Robotica sin dall'inizio del progetto Lab2Go, come scheda per il controllo dei motori e di alcuni sensori del robot.

A partire dall'a.s. 2020/21, la necessità di mostrare esperimenti a basso costo e facilmente ripetibili anche a casa o in laboratori poco attrezzati, hanno spinto gli organizzatori a dedicare grande rilievo all'utilizzo di *Arduino* anche per la Fisica. Grazie all'impegno della responsabile del progetto Prof.ssa Pia Astone, la sezione INFN di Roma ha, infatti, acquistato numerosi kit *Arduino* da fornire a tutor esterni e borsisti per mostrare a docenti e studenti esperienze che i ragazzi possono ripetere a costi molto contenuti (Fig. 8).

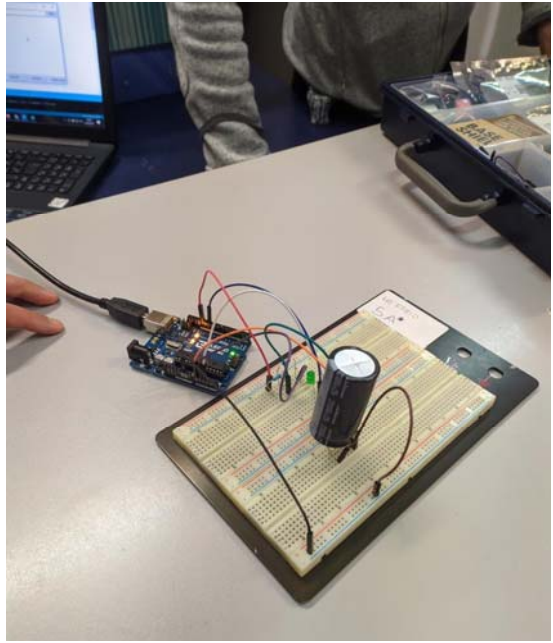

Figura 8: Esperimento sulla misura della costante di Planck con l'uso del kit *Arduino*, acquistato dalla sezione INFN di Roma, al Liceo Scientifico "*L. Trafelli*" di Nettuno (FR), a.s. 2020/21.

L'utilizzo di *Arduino* è in perfetta linea con lo spirito di *Lab2Go:* infatti, oltre a permettere di effettuare numerosi esperimenti grazie ai sensori che si possono facilmente acquistare a complemento della scheda, permette ai ragazzi di acquisire competenze trasversali come la capacità di programmare una macchina (*coding*) [9,12-15].

**2.3 La meccanica classica con i robot nell'ambito di Lab2Go-Robotica**

Nell'ambito della disciplina Robotica, l'uso del robot MARRtino viene applicato alla fisica, effettuando esperimenti di meccanica classica [16-17]. Infatti la verifica delle leggi fisiche studiate in classe (e in particolare di quelle del moto) costituisce una delle applicazioni dirette più intuitive in ambito didattico dell'utilizzo di MARRtino. Questa particolare attività ha richiesto agli studenti di Robotica un doppio impegno in termini scientifici: l'apprendimento di un linguaggio informatico utilizzato nel mondo della ricerca e l'inventiva di adeguare le possibilità di movimento del Robot al test di un fenomeno fisico riproducibile. L'utilizzo di MARRtino ha sicuramente stimolato la curiosità degli studenti, che si sono ingegnati sia nella rivisitazione di esperimenti classici, di importanza storica, come ad esempio il piano inclinato di Galileo (Fig. 9), sia nella costruzione di supporti esterni da applicare al Robot stesso per sperimentare fenomeni meccanici come l'attrito dinamico, la forza centrifuga o la composizione dei moti.

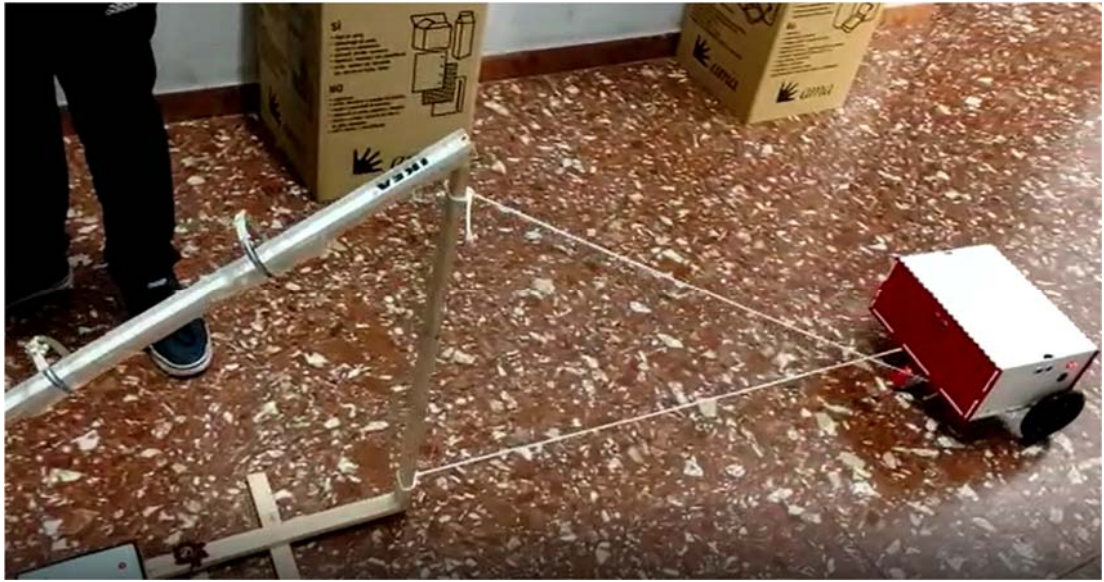

Figura 9: Robot *MARRtino* durante un esperimento sul piano inclinato in movimento presso il Liceo Scientifico "*Aristotele*" di Roma, a.s. 2020/21.

### 3. La WiKi di *Lab2Go*

Per poter catalogare gli strumenti presenti in laboratorio, le esperienze didattiche svolte e in generale le attività portate avanti dal progetto, la sezione di Roma dell'INFN ha realizzato una WiKi (Fig. 10), accessibile dal sito lab2go.roma1.infn.it/doku.php. Chiunque abbia accesso al web può consultare la WiKi e ne può, gratuitamente, utilizzare il materiale caricato; la scrittura delle pagine è, invece, possibile solo tramite account fornito dai responsabili del progetto, a tutors, borsisti, docenti e studenti delle scuole aderenti.

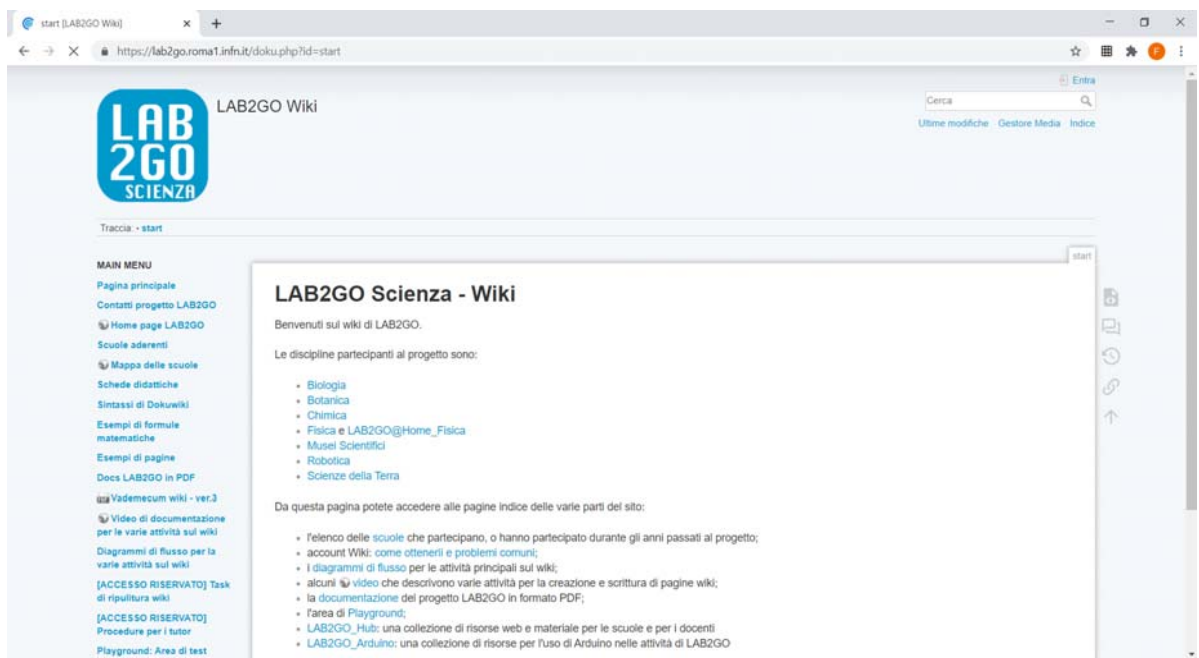

Figura 10: *Home-Page* della WiKi di *Lab2Go*.

La WiKi è suddivisa in sezioni in base alle discipline, cioè:
- Fisica;
- Lab2go@home;
- Musei scientifici;
- Arduino;
- Biologia Animale;
- Botanica;
- Chimica;
- Robotica;
- Scienze della Terra;

Per ciascuna sezione (namespace) sono creati sotto-namespace, per poter organizzare in modo efficiente i contenuti. Ad esempio, per la disciplina Fisica, le sotto-sezioni sono: strumenti, esperienze e schede didattiche (Fig. 11).

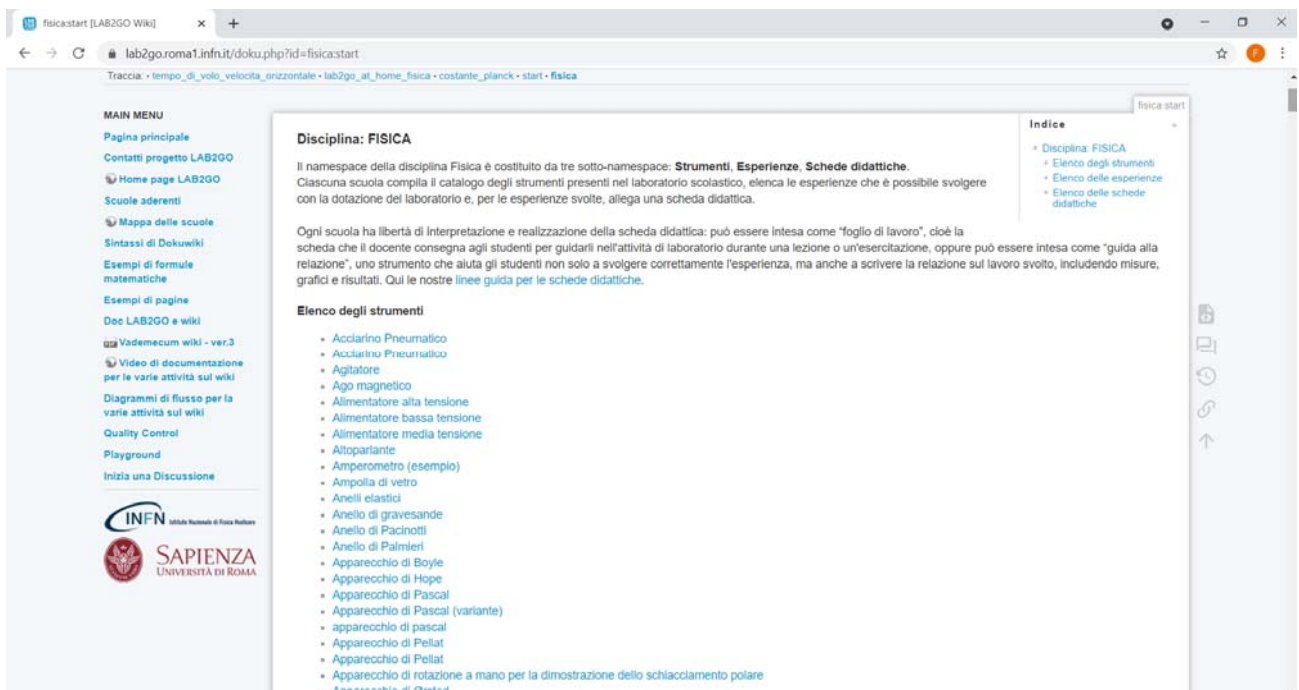

Figura 11: Sezione Fisica della WiKi di *Lab2Go*.

Alle sezioni sopra elencate, si aggiunge un'ulteriore sezione, chiamata *Playground*, in cui gli studenti possono esercitarsi ad impararne la sintassi.

### 3.1 La pagina personale della scuola

A ogni scuola partecipante è assegnata una pagina personale in cui gli studenti hanno il compito di catalogare, in delle apposite tabelle, gli strumenti o i materiali disponibili nel loro laboratorio e registrare le attività svolte nell'ambito del progetto. Nel caso della Fisica, una sezione-chiave è quella degli Strumenti, dove sono descritte le caratteristiche generali degli strumenti di laboratorio, documentate in un'area comune a tutte le scuole. Analogamente, anche le esperienze e le schede didattiche sono salvate in specifiche aree comuni della WiKi. Eventuali specifiche di uno strumento conservato in una determinata scuola sono, invece, descritte solo nella pagina personale della scuola stessa (Fig. 12).

Figura 12: Esempio di pagina personale del Liceo Classico *"P. Albertelli"* di Roma. Dalla prima colonna (Strumenti) è possibile accedere alle pagine di descrizione generali (salvate nell'area comune della WiKi) dei singoli strumenti catalogati. Dalla quarta colonna (Particolari Strumento Albertelli), invece, si accede alle caratteristiche specifiche dello strumento posseduto nella specifica scuola (tale descrizione è salvata nell'area privata della scuola).

**Conclusioni**
Tra i progetti PCTO organizzati dall'Università Sapienza di Roma per le scuole della regione Lazio, è di grande rilievo il progetto *Lab2Go*, rivolto agli studenti di scuole secondarie di II grado (dando loro la possibilità di avvicinarsi alla pratica laboratoriale) e che intende creare una rete di scuole che condividono pratiche laboratoriali con l'ausilio di Università ed Enti di Ricerca. Nato per la Fisica, in collaborazione con la sezione di Roma dell'INFN ed esteso poi ad altre discipline scientifiche, grazie alle ulteriori sezioni INFN che si sono successivamente interessate al progetto, *Lab2Go* si è diffuso anche in altre regioni. In seguito all'emergenza COVID-19, inoltre, è nato *Lab2Go@home* che promuove esperimenti fatti con materiali poveri o strumenti a basso costo quali app per *smartphone* e scheda *Arduino*. Gli strumenti utilizzati e le esperienze effettuate sono catalogati dagli studenti nella WiKi, che rappresenta uno strumento fondamentale per la condivisione di materiali, esperienze e pratiche didattiche, fornendo a tutti i docenti materiale facilmente consultabile sull'utilizzo dei laboratori didattici all'interno dei propri Istituti. Le testimonianze, finora raccolte tra docenti e studenti, hanno dimostrato che Lab2Go è stato un valido supporto ai docenti permettendo loro sia di approfondire argomenti trattati durante le ore di lezione, sia di introdurne altri non previsti dalla programmazione scolastica. Grazie a Lab2Go, gli studenti hanno imparato a realizzare un esperimento, raccogliere ed analizzare i dati, stimare gli errori sperimentali e compilare una scheda di laboratorio. Quanto detto riassume quanto fatto fino all'a.s. 2020/21, ma Lab2Go è un progetto tuttora attivo, per cui ci auguriamo che i docenti che leggeranno questo articolo vorranno proporre alla loro scuola, dal prossimo anno scolastico, la partecipazione al progetto.

**Ringraziamenti**

**Bibliografia**

[1] Lab2Go collaboration: web.infn.it/lab2go/ (Ultimo accesso: 19.11.2021)

[2] Lab2Go collaboration: lab2go.roma1.infn.it/doku.php (Ultimo accesso: 19.11.2021)

[3] G. Organtini et al.: Promoting the physics laboratory with Lab2Go- *Atti del 9th International Conference on Education and New Learning Technologies*, Barcellona, (2017)

[4] G. Organtini, R. Faccini: Esperienze innovative del PLS Roma- *Atti del 103th Congresso della Società Italiana di Fisica (SIF)*, Trento, (2017)

[5] M. Andreotti et al.: Lab2Go: A project for supporting laboratory practice in teaching STEM disciplines in high school- *Atti del 107th Congresso della Società Italiana di Fisica (SIF)*, On-line, (2021)

[6] M. Andreotti, et al.: Il progetto Lab2Go per la diffusione della pratica laboratoriale nell'insegnamento delle discipline STEM nelle scuole secondarie di II grado- *Atti del 59th Congresso dell' Associazione per l'insegnamento della Fisica (AIF)*, Roma, (2021)

[7] Docuwiki collaboration: www.dokuwiki.org/dokuwiki (Ultimo accesso: 19.11.2021)

[8] Marrtino collaboration: www.marrtino.org (Ultimo accesso: 19.11.2021)

[9] F. Casaburo: Teaching Physics by Arduino during COVID- 19 Pandemic: The Free Falling Body Experiment, *Physics Education,* vol. **56**, n. 6, p. 063001, https://doi.org/10.1088/1361-6552/ac1b39, (2021)

[10] Phyphox collaboration: phyphox.org (Ultimo accesso: 19.11.2021)

[11] Arduino collaboration: www.arduino.cc (Ultimo accesso: 19.11.2021)

[12] G. ORGANTNI: Arduino as a tool for physics experiments, *Journal of Physics: Conference Series*, vol. **1076**, p. 012026, (2018)

[13] G. ORGANTNI: "*Fisica con Arduino*" - Zanichelli, 2020

[14] G. ORGANTNI: "*Physics Experiments with Arduino and Smartphones*" - Springer, 2021

[15] F. CASABURO *et al.*:: Measurement of the Planck's constant in the framework of the Lab2Go project- *Atti del 107th Congresso della Società Italiana di Fisica (SIF)*, On-line, (2021)

[16] P. FERRARELLI, L. IOCCHI: Learning Newtonian Physics through Programming Robot Experiments, Tech Know Learn, doi.org/10.1007/s10758-021-09508-3, (2021)

[17] P. FERRARELLI, *et al.*: Improving Student's Concepts About Newtonian Mechanics Using Mobile Robots, in: Lepuschitz W et. al. (eds) *Robotics in Education. RiE 2018. Advances in Intelligent Systems and Computing*, vol **829.** Springer, Cham. doi.org/10.1007/978-3-319-97085-1_12, (2019)